\documentclass[dvips,aoas,preprint]{imsart}
\usepackage{amsmath,amssymb,amsthm,amsfonts,amstext,amsbsy,amscd}
\usepackage{graphicx}
\usepackage{multirow}
\RequirePackage{natbib}
\setattribute{journal}{name}{}

\usepackage{cite}
\usepackage[colorlinks=true,breaklinks=true]{hyperref}

\usepackage{framed}
\usepackage{ifthen}

\numberwithin{equation}{section}

\startlocaldefs

\newtheorem{theorem}{Theorem}[section]

\newtheorem{lemma}[theorem]{Lemma}

\newtheorem{definition}[theorem]{Definition}

\DeclareMathOperator{\E}{{\mathbb E}}
\DeclareMathOperator{\R}{{\mathbb R}}

\DeclareMathOperator{\MISE}{MISE}
\DeclareMathOperator{\rMISE}{rMISE}
\DeclareMathOperator{\supp}{supp}

\DeclareMathOperator{\Var}{Var} 
\DeclareMathOperator{\Cov}{Cov}
\renewcommand{\phi}{\varphi}

\newcommand{\MSE}{\operatorname{MSE}}
\newcommand{\SNR}{\operatorname{SNR}}
\newcommand{\cadlag}{c\`adl\`ag}
\newcommand{\SURE}{\operatorname{SURE}}
\newcommand{\cov}{\kappa}
\newcommand{\TV}{\operatorname{TV}}

\endlocaldefs

\begin{document}
\begin{frontmatter}
\title{Spot volatility estimation for high-frequency data: adaptive estimation in practice}
\runtitle{Spot volatility estimation in practice}

\begin{aug}
\author{\fnms{Till} \snm{Sabel}\corref{}\thanksref{m1,m3}\ead[label=e1]{tsabel@uni-goettingen.de}},
\author{\fnms{Johannes} \snm{Schmidt-Hieber}\thanksref{m2,m3}\ead[label=e2]{Johannes.Schmidt.Hieber@ensae.fr}}
\and
\author{\fnms{Axel} \snm{Munk}\thanksref{m1,m3}\ead[label=e3]{amunk1@gwdg.de}}

\thankstext{m3}{The research of the authors was supported by DFG/SNF-Grant FOR 916.}
\thankstext{m2}{The research of J. Schmidt-Hieber was funded by DFG postdoctoral fellowship SCHM 2807/1-1.}

\runauthor{T. Sabel et al.}

\affiliation{Georg-August-Universit\"at G\"ottingen\thanksmark{m1} and CREST-ENSAE\thanksmark{m2}}

\address{Institut f\"ur mathematische Stochastik\\Georg-August-Universit\"at G\"ottingen\\Goldschmidtstr. 7\\ 37077 G\"ottingen\\ Germany\\
\printead{e1}\\
\phantom{E-mail:\ }\printead*{e3}}

\address{\'Ecole Nationale de la Statistique\\ et de l'Administration \'Economique\\ Centre de Recherche\\ en \'Economie et Statistique\\ 3 Avenue Pierre Larousse\\ 92245 Malakoff\\ France\\
\printead{e2}}
\end{aug}

\begin{abstract}
We develop further the spot volatility estimator introduced in \citet{HoffmannMunkSchmidt-Hieber:2012} from a practical point of view and make it useful for the analysis of high-frequency financial data. In a first part, we adjust the estimator substantially in order to achieve good finite sample performance and to overcome difficulties arising from violations of the additive microstructure noise model (e.g. jumps, rounding errors). These modifications are justified by simulations. The second part is devoted to investigate the behavior of volatility in response to macroeconomic events. We give evidence that the spot volatility of Euro-BUND futures is considerably higher during press conferences of the European Central Bank. As an outlook, we present an estimator for the spot covolatility of two different prices.
\end{abstract}


\end{frontmatter}

\section{Introduction}
Semimartingales provide a natural class for modeling arbitrage-free log-price processes (cf. \citet{DelbaenSchachermayer:1994,DelbaenSchachermayer:1998}). In this context, estimation of the volatility and its surrogates such as integrated volatility or higher moments is inevitable for many purposes as for example hedging or option pricing. Under a semimartingale assumption, estimation of the volatility can be done using realized quadratic variation techniques (cf. for example \citet{FanWang:2008}). During the last decades, however, technical progress of trading platforms allowed to trade and to record data on very high frequencies. On these fine scales, microstructure effects due to market frictions have to be taken into account (for an overview of such market frictions cf. \citet{Hasbrouck:1993} and \citet{Madahavan:2000}). Following \citet{Zhou:1996}, these are often modelled by an additive noise process in the literature. Incorporating microstructure noise, our observations are given by
\begin{align}
	Y_{i,n} &=X_{i/n}+\epsilon_{i,n}, \quad i=1,\ldots,n
	\label{eq.mod_intro}
\end{align}
where the (latent) price process $X$ is considered to be a continuous It\^o semimartingale, that is $dX_t=\sigma_t dW_t+ $``drift``, with $W$ a Brownian motion. The quantity of interest, the volatility $\sigma$, has to satisfy some regularity conditions, in order to make everything well-defined. Adding the noise process $(\epsilon_{i,n})$ accounts for microstructure effects.

Microstructure noise leads to severe difficulties for estimation: As the noise is generally rougher than the original (latent) price process $X$, methods based on increments of the data become inconsistent as the resulting estimators are first order dominated by noise. For example, the realized quadratic variation does not converge to the integrated volatility as the sample size increases (cf. \citet{BandiRussell:2008}). Rather, it tends to infinity (cf. \citet{Zhou:1996}). See also \citet{Ait-SahaliaYu:2009} for a comprehensive empirical analysis of the noise level of different NYSE stocks.

Beginning with the work of \citet{Ait-SahaliaMyklandZhang:2005} and \citet{ZhangMyklandAit-Sahalia:2005}, various sophisticated regularization methods have been developed in order to estimate the integrated volatility under microstructure noise, cf. \citet{Zhang:2006}, \citet{FanWang:2007}, and \citet{Barndorff-NielsenHansenLundeStephard:2008}, to name just a few. Of particular interest in this work is the pre-average technique proposed in \citet{PodolskijVetter:2009} and \citet{JacodLiMyklandPodolskijVetter:2009}.

These methods target on integrated volatility, that is the spot volatility integrated over a fixed time interval. Estimation of the spot volatility, that is pathwise reconstruction of the volatility function $s \rightsquigarrow \sigma_s^2$ itself, has been less studied and is more complicated as it needs to combine tools from nonparametric statistics and stochastic analysis. Naive numerical differentiation of the integrated volatility does not perform well and additional regularization is required. In \citet{MunkSchmidt-Hieber:2010}, an estimator of the spot volatility was proposed, which is based on a Fourier series expansion of $\sigma^2$. Although this estimator could be shown to be asymptotically rate-optimal in Sobolev ellipsoids and hence is a first step towards a rigorous approach to spot volatility estimation, it suffers from various drawbacks. First, it obeys Gibb's effects which are well-known for Fourier estimators given non-smooth signals. Secondly, it requires knowledge of the smoothness of the 
underlying spot volatility, which is unknown in practice. To overcome these issues, \citet{HoffmannMunkSchmidt-Hieber:2012} introduced a wavelet estimator of $\sigma^2$. This estimator fully adapts to the smoothness of the underlying function and is rate-optimal over Besov classes. However, notice that Hoffmann et al. deals with the abstract estimation theory in model \eqref{eq.mod_intro} without making a particular connection to finance. We fill this gap in the current paper by specifically tuning the estimator for application to stock market data, while at the same time keeping the procedure purely data-driven and adaptive. In the following, we refer to the modified estimator as {\it Adaptive Spot Volatility Estimator (ASVE)}.

The key idea of the estimation method is to exploit the different smoothness properties of the semimartingale and the noise part: In a first step, we compute weighted local averages over data blocks of size $c\sqrt{n}$, for a constant $c>0$ independent of $n$. We show that the squared averages can be thought of as being observations in a regression type experiment. This is essentially the pre-averaging trick presented in \citet{JacodLiMyklandPodolskijVetter:2009} and \citet{PodolskijVetter:2009}. On one hand, local averaging reduces the impact of the noise (by a CLT type argument), while at the same time, the semimartingale part is (up to some small bias) not affected due to its a.s. H\"older continuity. On the other hand, treating the squared averages as new observations results in a reduction of the sample size from $n$ to $c^{-1}\sqrt{n}$. Pre-averaging might be also viewed as a denoising technique. In a second step, the pre-averaged data are decomposed via discrete wavelet transform and a robust 
thresholding procedure is applied. A detailed explanation concerning the construction of ASVE is given in Section \ref{sec:est}.

Let us summarize in the following the main difficulties that we address in order to make the estimator applicable to real financial data.

\begin{enumerate}
 \item {\it Thresholding:} One of the main challenges is to find a suitable and robust wavelet thresholding method. We argue in Section \ref{sec:heuristic} that rewriting the initial model via the pre-average transform yields, as outlined above, a regression model with errors following approximately a centered $\chi_1^2$-distribution. Furthermore, the errors are dependent and heteroscedastic causing severe difficulties for wavelet estimation. Therefore, a crucial point in our method is the choice of the thresholding procedure. We address this problem in Section \ref{sec:thresholding}.
 \item {\it Parameter tuning:} ASVE requires to pick a bandwidth and a weight function. The specific choice will heavily influence the finite sample performance and even the asymptotic variance. In Section \ref{sec:opttuning}, we propose a method to chose these values based on an explicit computation of the asymptotic variance in a toy model. In a second part, the finite sample performance for these choices is studied in simulations.
 \item {\it Model violations:} Given real data, model violations often occur. These include rounding errors, which is a non-additive microstructure effect as well as various types of jumps (cf. \citet{Ait-SahaliaJacod:2009}, \citet{Ait-SahaliaJacodLi:2012}). In Section \ref{sec:jumpsim}, we show that rounding has almost no impact on the performance of the estimator, while the presence of jumps is indeed a very delicate problem. In order to eliminate jumps in the price, we propose in Section \ref{sec:jumptest} a specific pre-processing of the data.
 \item {\it Trading times:} We have to deal with data recorded at non-equi\-dis\-tant time points. One possibility to 'convert' data into the equispaced framework of model \eqref{eq.mod_intro} is to subsample the process, that is to sample for example every 10th second. In Section \ref{sec:ticktime}, we propose another method by defining different time schemes. Especially, we distinguish between real time and tick time and clarify their connection.
\end{enumerate}

While Section \ref{sec:calibration} is devoted to calibration of ASVE especially focussing on the issues mentioned above, in Section \ref{sec:sim}, we evaluate ASVE by numerical simulations. This includes a stability analysis regarding model violations and different types of microstructure noise.

\begin{figure}[t!]
 \includegraphics[width=\textwidth]{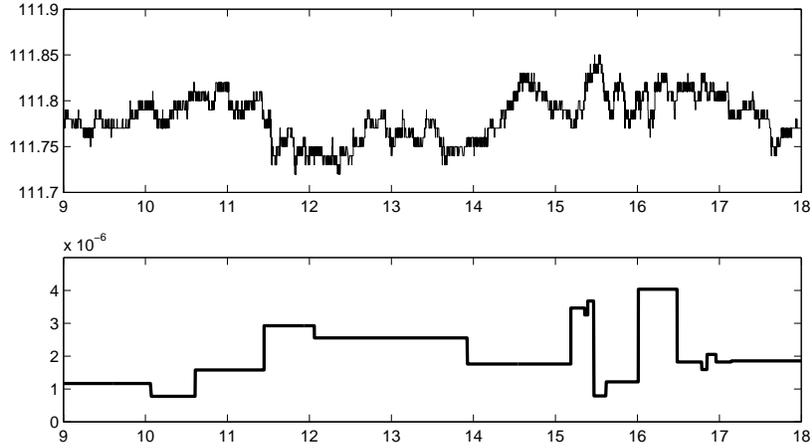}
\caption{Application to real data. Upper panel: Price of FGBL data on June, 4th 2007. Lower panel: Adaptive spot volatility estimator (ASVE).}\label{fig:intro}
\end{figure}

As an illustrating example for a real data application, Figure \ref{fig:intro} shows Euro-BUND (FGBL) prices for June 4th, 2007 together with the reconstructed volatility.
Notice that ASVE appears to be locally constant which is due to the specific wavelets which are the building blocks of this estimator. Note further, that ASVE is still quite regular, while spot volatility is commonly assumed to have no finite total variation. This relies on the fact that microstructure noise induces additional ill-posedness to the problem which leads to relatively slow convergence for any estimator (cf. \citet{Reiss:2011}). Therefore, only key features of the spot volatility can be expected to be reconstructed, while fine details cannot be recovered by any method.

In Section \ref{sec:real}, a more extensive investigation of real data is done concerning the reaction of spot volatility in answer to macroeconomic announcements: We study characteristics of the volatility of FGBL prices during the monthly ECB press conference on key interest rates. We observe that the spot volatility as well as the volatility of the volatility is higher during these conferences.

Finally in Section \ref{sec:multi}, we discuss extensions of ASVE to spot covolatility estimation.

\section{The Adaptive Spot Volatility Estimator (ASVE)}\label{sec:est}

\subsection{Wavelet estimation} A common tool for adaptive, nonparametric function estimation is wavelet thresholding (cf. for example \citet{DonohoJohnstone:1994} and \citet{DonohoJohnstoneKerkyacharianPicard:1995}, for some early references). Assume our signal, say $f$, is a function in $L^2[0,1]$. Then, for given scaling function $\phi$ and corresponding wavelet $\psi$, the function $f$ can be decomposed into
\begin{align}
    f=\sum_{k} \left\langle f, \phi_{j_0,k} \right\rangle \phi_{j_0,k}
    +\sum_{j=j_0}^\infty \sum_{k \in \mathbb{Z}} \left\langle f, \psi_{j,k} \right\rangle \psi_{j,k}, \quad j_0\in \mathbb{N},
    \label{eq.wavexp}
\end{align}
where the sum converges in $L^2[0,1]$. Here, $\langle f,g\rangle= \int_0^1 f(x)g(x)dx$, $\phi_{j,k}(\cdot)=2^{j/2}\phi(2^j\cdot-k)$, and $\psi_{j,k}(\cdot)=2^{j/2}\psi(2^j\cdot-k)$. The scaling and wavelet coefficients are $\left\langle f, \phi_{j_0,k} \right\rangle$ and $\left\langle f, \psi_{j,k} \right\rangle$, respectively. See \citet{Daubechies:1992}, \citet{Cohen:2003} for an introduction to wavelets, \citet{CohenDaubechiesVial:1993} for wavelets on $[0,1]$, and \citet{Wassermann:2010} for a reference to wavelets in statistics.

Suppose that we have estimators for scaling and wavelet coefficients, denoted by $\widehat{\left\langle f, \phi_{j_0,k} \right\rangle}$ and $\widehat{\left\langle f, \psi_{j,k}\right\rangle}$, respectively. A thresholding estimator for $f$ is given by
\begin{align}
    \widehat f = 
    \sum_{k} \widehat{\langle f, \phi_{j_0,k}\rangle} \phi_{j_0,k}
    +\sum_{j=j_0}^{j_1} \sum_{k \in \mathbb{Z}} \mathcal{T}\big(\widehat{\langle f, \psi_{j,k}\rangle}\big) \psi_{j,k},
    \label{eq.waveest}
\end{align}
for some thresholding procedure $\mathcal{T}$. Traditional choices for $\mathcal{T}$ include hard thresholding ($\mathcal{T}_{HT}(x)=x\mathbf{1}_{\{|x|>t^\ast\}}$) and soft thresholding ($\mathcal{T}_{ST}(x)=(x-t^\ast)\mathbf{1}_{\{x>t^\ast\}}+(x+t^\ast)\mathbf{1}_{\{x<-t^\ast\}}$), both for some threshold level $t^\ast$. The idea of term-by-term thresholding is to keep large coefficients while discarding small ones for which one cannot be sure that they contain significant information about the true signal. 

Even though coefficientwise thresholding has many appealing theoretical properties, it nevertheless might lead to unstable reconstructions if applied to real data. Robustification of wavelet thresholding is typically based on variations of the following idea. Assume for the moment that $\psi$ is the Haar wavelet, which has compact support on $[0,1]$. Then, $\left\langle f, \psi_{j,k} \right\rangle$ depends only on $f$ restricted to the interval $[2^{-j}k,2^{-j}(k+1)]$. If the absolute value of the estimate of $\left\langle  f, \psi_{j,k} \right\rangle$ is large, while the absolute values of the estimates of nearby coefficients are small, then it is likely that this is due to an outlier and hence the wavelet coefficient should be discarded as well.

There are two types of methods for detecting such situations. Tree-struc\-tured wavelet thresholding using the hierarchical pattern of multiresolution analysis (cf. for example \citet{AutinFreyermuthSachs:2011}) and block thresholding methods, which are based on neighboring coefficients for fixed level $j$. For our problem, SURE block thresholding (cf. \citet{CaiZhou:2009}) turns out to work well. For more details, we refer to Section \ref{sec:thresholding} as well as Section \ref{sec:sim}.

\subsection{Model}

Consider the process $X$ defined via $dX_t=\sigma_t dW_t+ b_tdt$ and $X_0=0$ on a filtered probability space $(\Omega, \mathcal{F},(\mathcal{F}_t)_{t\geq 0},\mathbb{P})$, where $W$ denotes a standard Brownian motion. The processes $\sigma$ and $b$ are assumed to be $\mathcal{F}_t$-adapted and \cadlag. We will always suppose that $\sigma$ and $b$ are chosen in such a way that a unique weak solution of the SDE above exists. 

Recall \eqref{eq.mod_intro}, that is we observe 
\begin{align}
	Y_{i,n} &=X_{i/n}+\epsilon_{i,n}, \quad i=1,\ldots,n.
	\label{eq.mod}
\end{align}
While $X$ should be interpreted as the true, uncorrupted price process, the noise process $(\epsilon_{i,n})$ models the microstructure effects. We allow for inhomogeneous variation in the noise, that is
\begin{align}
	\epsilon_{i,n}=\tau(\tfrac in, X_{i/n}) \eta_{i,n},\label{eq.noise}
\end{align}
where $(\eta_{i,n})_i$ is an i.i.d. sequence of random variables independent of $X$. Notice that the noise level may depend on the price itself. For identifiability, we assume further that $(\eta_{i,n})_i$ is centered and second moment normalized, that is $\E\eta_{i,n}^2=1$ for $i=1,\ldots,n$.

To summarize, $Y_{i,n}$ is the observed price, which is the sum of the latent true price process $X$ at time point $i/n$ under additional microstructure noise $\epsilon_{i,n}$.

While the drift $b$ is of only minor importance for high-frequency data, the volatility $\sigma$ is the key quantity in this model as it drives the fluctuation and variation behavior of the process. Although under debate, the additive microstructure noise model \eqref{eq.mod} is commonly believed to perform very well in practice, as it is able to reproduce many stylized facts found in empirical financial data. Moreover, to the best of our knowledge, it is the only model incorporating microstructure noise for which a theory of pathwise estimation of the volatility exists.

\subsection{Pre-averaging and estimation of series coefficients}
 The key step behind the construction of ASVE is a transformation of the data, which allows to rewrite the original problem as a nonparametric regression problem. This transformation is based on the pre-averaging method as introduced in \citet{JacodLiMyklandPodolskijVetter:2009} and \citet{PodolskijVetter:2009}. Since then, pre-averaging became an important tool to tackle estimation under microstructure. For an extension of pre-averaging to data measured on an endogenous time grid, cf. \citet{LiZhangZheng:2013}. Recently, the practical performance of these methods in estimation of integrated volatility was investigated in \citet{HautschPodolskij:2013}.

In a first step, let us introduce a class of suitable weight functions (cf. Hoffmann et al., Definition 3.1).

\begin{definition}[Pre-average function]
\label{pre-average function}
A piecewise Lipschitz continuous function $\lambda : [0,2]\rightarrow \R$ satisfying $\lambda(t)=-\lambda(2-t)$, for all $t\in [0,1]$ and 
\begin{equation}
  \Big(2\int_{0}^1 \big(\int_0^s \lambda(u)du\big)^2 ds\Big)^{1/2}=1
  \label{eq.norm_condition_preav}
\end{equation}
is called a (normalized) pre-average function. 
\end{definition}

Notice that whenever we have a function $\widetilde \lambda$ satisfying all assumptions of the previous definition except \eqref{eq.norm_condition_preav}, then by dividing $\widetilde \lambda$ through the l.h.s. of \eqref{eq.norm_condition_preav}, we obtain a proper pre-average function. Next, we define local averages using weights generated from pre-average functions.

Define $m=n/\lfloor n^{1/2}/c \rfloor$ for some fixed $c>0$. Notice that $m=c\sqrt{n}+O(1)$ and that $m$ divides $n$. The divisibility property allows to get rid of some discretization errors later. For $i=2,...,m$, set
\begin{align}\label{eq.ovYdef}
 \overline Y_{i,m}(\lambda):=\frac mn \sum_{\frac jn\in[\frac{i-2}m,\frac{i}m]}\lambda\big(m\tfrac
jn-(i-2)\big)Y_{j,n}.
\end{align}
Further, let us introduce
\begin{align*}
\mathfrak{b}(\lambda, Y_\cdot)_{i,m} 
:=  &\frac{m^2}{2n^2} \sum_{\frac jn\in[\frac{i-2}m,\frac im]}\lambda^2\big(m\tfrac jn-(i-2)\big)\big(Y_{j,n}-Y_{j-1,n}\big)^2
\end{align*}
which plays the role of a bias correction. For any $L^2$-function $g$, the estimator of the scalar product $\langle  \sigma^2, g\rangle$ is given by its empirical version applied to the bias-corrected squares $\overline{Y}_{i,m}^2$ via
\begin{equation} \label{def est fonct lin}
\widehat{\langle \sigma^2, g\rangle} :=\sum_{i=2}^m g\big(\tfrac{i-1}{m}\big)\big[ \ \overline{Y}_{i,m}^2-\mathfrak{b}(\lambda,Y_\cdot)_{i,m}\big]
  =\frac 1m \sum_{i=2}^m g\big(\tfrac{i-1}{m}\big) Z_{i,m},
\end{equation}
where 
\begin{equation}
  Z_{i,m} := m \big[ \ \overline{Y}_{i,m}^2-\mathfrak{b}(\lambda,Y_\cdot)_{i,m}\big].
  \label{eq.pre_average_def_eq}
\end{equation}

\begin{definition}
\label{definition.preaverage values}
 The random variables $Z_{i,m}$, $i=1,\ldots,m$, are called {\it pre-averaged values}.
\end{definition}

As we will show below, the pre-averaged values can be interpreted as observations coming from a nonparametric regression experiment with the spot volatility being the regression function. For $g\in \{ \phi_{j_0,k}, \psi_{j,k} \}$, we obtain estimates for the scaling/wavelet coefficients $\langle \sigma^2, \phi_{j_0,k}\rangle$ and $\langle \sigma^2, \psi_{j,k}\rangle$, respectively. In practice, fast computations of these coefficients can be performed using a discrete wavelet transform (DWT).

\subsection{A heuristic explanation}\label{sec:heuristic} In this part, we will present the main idea underlying the construction of the estimator, which is to think of the pre-averaged values $(Z_{i,m})_i$ as coming from a nonparametric regression problem. First, note that for $i=2,\ldots,m$,
\begin{align*}
\overline Y_{i,m}(\lambda)\approx \int_{\tfrac{i-2}m}^{\tfrac im} m \lambda\big(ms-(i-2)\big)X_sds  + \xi_{i,m}
\end{align*}
with 
\begin{align*}
  \xi_{i,m}
=\frac mn \sum_{\frac jn\in[\frac{i-2}m,\frac im]}\lambda\big(m\tfrac
jn-(i-2)\big)\epsilon_{j,n}.
\end{align*}
Now, let $\Lambda(u)=-\int_0^u  \lambda(v)dv \mathbb{I}_{[0,2]}(u)$. By Definition \ref{pre-average function}, $\Lambda(0)=\Lambda(2)=0$. Hence, $\Lambda'(ms-(i-2))=m \lambda(ms-(i-2))$ and using partial integration
\begin{align*}
  \overline{Y}_{i,m}
  &\approx \int_{\tfrac{i-2}m}^{\tfrac im}
  \Lambda\big(ms-(i-2)\big)dX_s+ \xi_{i,m}.
\end{align*}
It is easy to verify that $\xi_{i,m}=O_p(\sqrt{m/n})$ and $\E\xi_{i,m}^2=\E\mathfrak{b}(\lambda,\epsilon_\cdot)_{i,m}\approx \E\mathfrak{b}(\lambda,Y_\cdot)_{i,m}$. For the diffusion term, $\int_{(i-2)/m}^{i/m} \Lambda(ms-(i-2))dX_s=O_p(m^{-1/2})$ and by It\^o's formula there exists $U_{i,m}$, such that $\E U_{i,m}=0, \  U_{i,m}=O_P(m^{-1})$, and
\begin{align*}
\begin{aligned}
  \Big(\int_{\tfrac{i-2}m}^{\tfrac im}
  \Lambda\big(ms-(i-2)\big)dX_s\Big)^2 
  &=
  \int_{\tfrac{i-2}m}^{\tfrac im}\Lambda^2\big(ms-(i-2)\big)\sigma_s^2 ds+ U_{i,m}\\
  &\approx \frac 1m\sigma^2_{(i-1)/m}+U_{i,m},
\end{aligned}
\end{align*}
using the definition of a pre-average function for the last step. Recall \eqref{eq.pre_average_def_eq}. Then, $\E[ Z_{i,m}-\sigma_{(i-1)/m}^2]\approx 0$ and $Z_{i,m} -\sigma_{(i-1)/m}^2  =O_P\Big(1+ \frac{m}{n^{1/2}}+\frac{m^2}n\Big)=O_P(1)$, since $m= c\sqrt{n}+O(1)$. To summarize,
\begin{align}
  Z_{i,m}=\sigma_{(i-1)/m}^2+\widetilde \epsilon_{i,m}, \quad i=2,\ldots,m,
  \label{eq.regression_form}
\end{align}
with $\E\widetilde \epsilon_{i,m}\approx 0$ and $\widetilde \epsilon_{i,m}=O_P(1)$. Hence, we may interpret $(Z_{i,m})_{i=2,\ldots,m}$ as a random vector generated from a regression problem with regression function $\sigma^2$ and additive (dependent) noise $\widetilde \epsilon_{i,m}$. 

Let us conclude this section with the following remarks.

\begin{itemize}
  \item[-] Notice that the estimator of $\widehat{\langle \sigma^2, g \rangle}$ in \eqref{def est fonct lin} is just the empirical version of the scalar product $\langle \sigma^2, g\rangle$ in the regression model \eqref{eq.regression_form}. 
  \item[-] By some CLT argument, the distribution of $\overline Y_{i,m}$ as defined in \eqref{eq.ovYdef}, will converge to a Gaussian law. But since we are considering the squares of $\overline{Y}_{i,m}$ in \eqref{eq.pre_average_def_eq}, the noise process in \eqref{eq.regression_form} will not be Gaussian. Rather, one can think of the $\widetilde\epsilon_{i,m}$'s as centered $\chi_1^2$ random variables.
 \item[-] The variance of $\widetilde \epsilon_{i,m}$ (which is here approximately the second moment) is (up to some remainder terms) a quadratic function in $\sigma_{i/n}$ and $\tau(i/n, X_{i/n})$. Therefore, the regression problem \eqref{eq.regression_form} is strongly heteroscedastic. This point is separately addressed in Section \ref{sec:thresholding}.
 \item[-] Rewriting the original problem as regression model, as outlined above, reduces the effective number of observation from $n$ to $m$ and thus to the order $n^{1/2}$. This implies that if we can estimate a quantity in the regression model (for example pointwise estimation of the regression function $\sigma^2$) with rate $m^{-s}$, given $m$ observation, we obtain the rate of convergence $n^{-s/2}$ in the original model \eqref{eq.mod}. Therefore, we always lose a factor $1/2$ in the exponent of the rate of convergence. It is well-known that this is inherent to spot volatility estimation under microstructure noise. As proved in \citet{MunkSchmidt-Hieber:2010}, \citet{Reiss:2011} for various situations, these rates are optimal.
\end{itemize}

\subsection{Thresholding and construction of ASVE}\label{sec:thresholding}
Having the estimates of the wavelet coefficients at hand, let us outline the thresholding procedure. The proposed method extends SURE block thresholding as introduced in \citet{CaiZhou:2009} to heteroscedastic problems.

In order to formulate the thresholding estimator define, for a vector $v$, Stein's unbiased risk estimate (SURE) as
\begin{align*}
 \SURE(v,\lambda,L)= L+ \frac{\lambda^2 -2\lambda(L-2)}{\|v\|_2^2}
  \mathbb{I}_{\{\|v\|_2^2> \lambda\}}
  +(\|v\|_2^2-2L) \mathbb{I}_{\{\|v\|_2^2\leq \lambda\}}.
\end{align*}

First, we start with SURE block thresholding for homoscedastic data. For convenience, set $\widehat d_{j,k}= \widehat{\langle \sigma^2, \psi_{j,k}\rangle}$.

\begin{itemize}
 \item[In:]$j_0, j_1$, $(\widehat d_{j,k})_{j_0\leq j\leq j_1, k}$
 \item[(A)] For every fixed resolution level $j_0\leq j\leq j_1$ define $D_j$ as the set of wavelet dilations $\{k: k\in \mathbb{Z}, \ [0,1]\cap \supp \psi_{j,k} \neq \varnothing\}$. Denote by $T_{j}$ the mean of the random variables $\{(\widehat d_{j,k})^2-1: k\in D_j\}$ and consider the threshold $\gamma(u)=u^{-1/2}\log_2^{3/2}(u)$.
 \item[(B)] For any given vector $v \in \mathbb{R}^d$ and positive integer $L$ define the $q$th block (of length $L$) as $v^{(q,L)}=(v_{(q-1)L+1},\ldots, v_{qL \wedge d})$, $q\leq d/L$. Let $d=|D_j|$. In particular, denote by $(\widehat d_{j,k})_{k\in D_j}^{(q,L)}$ the $q$th block of length $L$ of the vector $(\widehat d_{j,k})_{k\in D_j}$ and define
 \begin{align*}
    (\lambda^\star, L^\star)
    =
    \arg\min_{\substack{1\leq L\leq d^{1/2}\\(L-2)\vee 0 \leq \lambda \leq 2L\log d}}
    \sum_{q=1}^{\lfloor d/L\rfloor} \SURE\big((\widehat d_{j,k})_{k\in D_j}^{(q,L)}, \lambda, L\big),
 \end{align*}
where $\lfloor . \rfloor$ is the floor function.
 \item[(C)] For every $k\in D_j$, the block thresholded (and standardized) wavelet coefficient is given by 
 \begin{align*}
    \mathcal{T}(\widehat d_{j,k})
    &=
    \begin{cases}
      (1-(2\log d) \ \widehat d_{j,k}^{-2})_+ \ \widehat d_{j,k}, 
      \quad &\text{if} \ T_j \leq \gamma(d), \\
      \big(1-\lambda^\star \big\|(\widehat d_{j,\ell})_{\ell \in D_j}^{(q(k), L^\star)}\big\|_2^{-2}\big)_+\ \widehat d_{j,k},
      &\text{if}  \ T_j > \gamma(d),
    \end{cases}
 \end{align*}
with $q(k)$ the (unique) block of length $L^\star$ including $k$. 
\item[Out:]$\mathcal{T}(\widehat d_{j,k})_{j_0\leq j\leq j_1, k}$.
\end{itemize}

SURE block thresholding optimizes levelwise over the block size $L$ and the shrinkage parameter $\lambda$ in step (B). However, it is well-known that this method does not yield good reconstructions in the case where only a few large wavelet coefficients are present. In order to circumvent these problems, in step (C), soft shrinkage is applied if $T_j$ is small.

As an additional difficulty, we have to deal with errors in \eqref{eq.regression_form}, that are heteroscedastic with unknown variance. Therefore, we normalize the wavelet coefficients by its standard deviation in a first step, that is for sets $I_{j,k}$, chosen below, define the empirical standard deviation on $I_{j,k}$ by
\begin{align}
  \widehat s_{j,k}
  := \Big[
  \frac{1}{|I_{j,k}|-1}\sum_{\tfrac im \in I_{j,k}}
  \Big( Z_{i,m}- \tfrac 1{|I_{j,k}|}\sum_{\tfrac im \in I_{j,k}} Z_{i,m}\Big)^2 
  \Big]^{1/2}
  \label{eq.def_std_dev}
\end{align}
and the {\it standardized wavelet coefficients} by $\widetilde d_{j,k}:= \widehat d_{j,k}/\widehat s_{j,k}$. Now, we run the SURE algorithm applied to $(\widetilde d_{j,k})_{j_0\leq j\leq j_1, k}$ instead of $(\widehat d_{j,k})_{j_0\leq j\leq j_1, k}$. In a final step we need to invert the standardization. Thus, the thresholded wavelet coefficients are given by $(\widehat s_{j,k} \mathcal{T}(\widetilde d_{j,k}))_{j_0\leq j\leq j_1, k}$. Together with the (truncated) series expansion \eqref{eq.wavexp}, we have

\begin{definition}ASVE is defined by
\begin{align*}
    \widehat \sigma^2(t)=\sum_{k}\widehat{\langle \sigma^2, \phi_{j_0,k} \rangle} \phi_{j_0,k}(t)
    +\sum_{j=j_0}^{j_1} \sum_{k \in D_j} \widehat s_{j,k} \mathcal{T}(\widetilde d_{j,k}) \psi_{j,k}(t), \quad t\in[0,1].
\end{align*}
\end{definition}

For estimation of the standard deviations $\widehat s_{j,k}$, one would instead of \eqref{eq.def_std_dev} rather prefer a robust estimate based on the median (cf. \citet{CaiZhou:2009}, p. 566) or to use variance stabilizing transformations. Since the error variables $\overline \epsilon_{i,m}$ in \eqref{eq.regression_form} do not follow a certain prespecified distribution, these approaches are not easily applicable here. Therefore, we rely on \eqref{eq.def_std_dev} and robustify our estimates by the choice of $I_{j,k}$, as described in the next paragraph: 

We pick some $j_I$, $j_0\leq j_I\leq j_1$. If $j\leq j_I$, we define $I_{j,k}$ as the support of $\psi_{j,k}$. For high resolution levels $j>j_I$, we enlarge the support of $\psi_{j,k}$ such that the length of $I_{j,k}$ never falls below $2^{-j_I}$. This guarantees some minimal robustness of the method.

Block thresholding uses the normality of the wavelet coefficients at various places. Thus, to ensure good performance, we need to check whether the distribution of the estimated wavelet coefficients follow approximately a Gaussian law. This is not obvious, because, as we argued in Section \ref{sec:heuristic}, the errors in the regression model \eqref{eq.regression_form} behave like centered $\chi_1^2$ random variables. However, since the estimator \eqref{def est fonct lin} is a weighted average of the observations, we indeed find 'almost' Gaussian wavelet coefficients in simulations. Thus, we do not need to include a further correction to account for the non-Gaussianity. Notice that these issues are closely linked to nonparametric variance estimation (cf. \citet{CaiWang:2008}). 

\section{Calibration and robustness}\label{sec:calibration}
\subsection{Optimal tuning parameters}\label{sec:opttuning}

In this section we propose empirical rules for choosing some variables in the ASVE procedure. Notice that the method requires to pick a pre-average function $\lambda$ and a constant $c>0$ defining the number of blocks $m$. By computing the asymptotic variance of ASVE in a simplified model, we derive some insight which pre-average functions might work well. In particular, this shows that $\lambda$ and $c$ should be chosen dependent on each other, that is $c=c(\lambda)$. In a second step, we study the finite sample performance of these choices for simulated data.

We start with investigating different choices for $\lambda$ and $c=c(\lambda)$ in a simplified version of model \eqref{eq.mod} for which the leading term of the mean squared error can be calculated explicitly.
\begin{lemma}\label{lem:mse} Work in model \eqref{eq.mod} with constant $\sigma, \tau$ and $\eta_{i,n}\sim \mathcal{N}(0,1)$ i.i.d. Then,
\begin{align*}
  \MSE(\widehat{\langle \sigma^2, 1\rangle})
  &= 
  \frac{4}{c}\Big(\int_0^1\sigma^2\Lambda(u)\Lambda(1-u)-(\tau c)^2 \lambda(u) \lambda(1-u)du
   \Big)^2n^{-1/2}
  \\
  &\quad\quad\quad\quad +\frac{2}{c}\Big(\sigma^2+2(\tau c)^2\|\lambda\|_{L^2[0,1]}^2\Big)^2n^{-1/2}+o(n^{-1/2}).
\end{align*}
\end{lemma}
A proof of this lemma can be found in \citet{Schmidt-Hieber:2010}, Section 5.4. Given a pre-average function $\lambda$, it allows us to compute the corresponding optimal constant $c^\star$ by minimizing the asymptotic $\MSE$. In general $c^\star$ is a multiple of the signal-to-noise ratio ($\SNR$), that is  $c^\star=$const$.\times \tfrac{\sigma}{\tau}$, where the constant depends on $\lambda$. In Table \ref{tab:preav}, the value of this constant for different pre-average functions and the leading term for the corresponding $\MSE$ are derived. 

\begin{table}[hb!]
\begin{center}
    \renewcommand{\arraystretch}{1.5}
\begin{tabular*}{\textwidth}{@{\extracolsep{\fill}}rccc}
\hline
\, $i$&$\lambda_i(s)=$&$c^\star\tau/\sigma\approx$&$\lim_nn^{1/2}(\tau\sigma^3)^{-1}\cdot\MSE\approx$\\
\hline
1&$\tfrac\pi2\cos(\tfrac\pi2s)$&0.49&10.21\\
2&$\tfrac{3\pi}2\cos(\tfrac{3\pi}2s)$&0.17&31.36\\
3&$\sqrt{\tfrac32}(\mathbb{I}_{[0,1)}(s)-\mathbb{I}_{(1,2]}(s))$&0.35&10.74\\
4&$\tfrac{\pi}{\sqrt{3}}\sin(\pi s)$&0.30&12.52\\
5&$\tfrac{2\pi}{\sqrt{3}}\sin(2\pi s)$&0.19&24.35\\
6&$\tfrac{3\sqrt{5}}2(1-s)^3$&0.47&20.41\\
7&$\tfrac{\sqrt{91}}2(1-s)^5$&0.38&20.36\\
\hline
\end{tabular*}
\caption{Different choices for pre-average functions, the optimal tuning parameter $c^\star$ as well as the asymptotic constant of the $\MSE$ for estimation of the integrated volatility.}
\label{tab:preav} 
\end{center}
\end{table}
\renewcommand{\arraystretch}{1}

It is well-known (cf. \citet{GloterJacod:2001, GloterJacod:2001a}, \citet{CaiMunkSchmidt-Hieber:2010}) that $\MSE(\widehat{\langle \sigma^2, 1\rangle})=8\tau\sigma^3n^{-1/2}(1+o(1))$ is asymptotically sharp in minimax sense. However, this minimum cannot be achieved within the class of estimators introduced in Section \ref{sec:est}. Using calculus of variations, we find that the best possible choice for the simplified model introduced above is $\lambda(\cdot)= \pi \cos(\cdot\pi/2)/2$. According to Table \ref{tab:preav}, the corresponding $\MSE$ is $10.21 \tau\sigma^3n^{-1/2}(1+o(1))$ achieving the optimal variance $8\tau\sigma^3n^{-1/2}(1+o(1))$  up to a factor $1.27$.

Computation of $c^\star$ requires knowledge of the $\SNR$, that is $\sigma/\tau$. As this is unknown, we suggest to estimate the $\SNR$ in a first step from the data via
\begin{align}
  \widehat\SNR = \left(\frac{\widetilde{\langle \sigma^2, 1\rangle}}{\widehat{\langle  \tau^2, 1\rangle}}\right)^{1/2},
  \label{eq.cestdef}
\end{align}
with rescaled quadratic variation $\widehat{\langle \tau^2, 1\rangle}=(2n)^{-1}\sum_{i=2}^n(Y_{i,n}-Y_{i-1,n})^2$  and 
\begin{align*}
  \widetilde{\langle \sigma^2, 1\rangle}
  := 
  \sum_{i=2}^{\widetilde m}
  \big(\overline Y_{i,\widetilde m}^2-\mathfrak{b}(\lambda,Y_\cdot)_{i,\widetilde m}
  \big), \quad \text{with} \ \widetilde m=\lfloor n^{1/2} \rfloor
\end{align*} 
as preliminary estimator of $\langle \sigma^2, 1\rangle$. It is easy to show that $\widehat{\langle \tau^2, 1\rangle}$ is $n^{1/2}$-consistent for estimation of the integrated noise level $\langle \tau^2, 1\rangle$ and since we are interested in data sets with sample size $n\sim 10^5$, we may directly divide by $\widehat{\langle \tau^2, 1\rangle}$ in \eqref{eq.cestdef} without any further regularization.

In the second part of this section, we study the finite sample performance for different pre-average functions. As Table \ref{tab:preav} suggests, the $\MSE$ deteriorates if the number of oscillations of $\lambda$ increases. Therefore, we choose the functions $\lambda_1(\cdot):= \pi \cos(\cdot\pi/2)/2$ (the optimal pre-average function in the simplified model), $\lambda_3(\cdot):= (\frac32)^{1/2}(\mathbb{I}_{[0,1)}-\mathbb{I}_{(1,2]})$ (the pre-average function used in \citet{HautschPodolskij:2013}), and $\lambda_4(\cdot):= \pi \sin(\cdot\pi)/3^{1/2}$ as possible candidates. 

\begin{figure}[h!]
 \includegraphics[width=\textwidth]{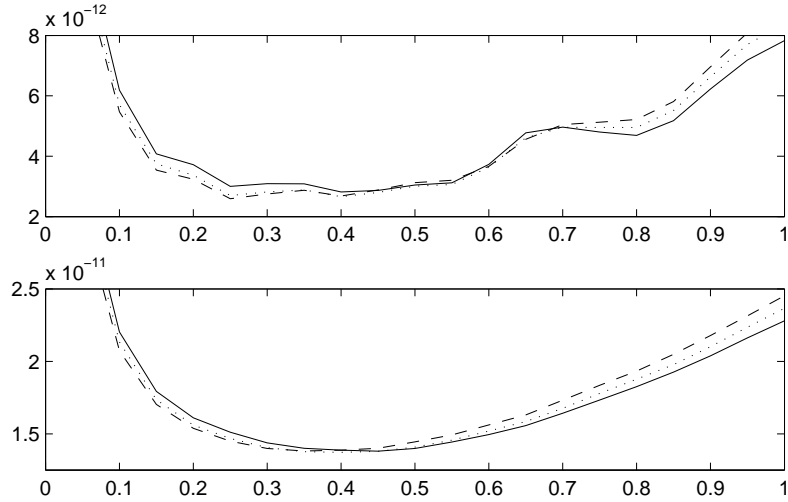}
\caption{Empirical MISE for 10,000 repetitions and data with constant $\sigma^2\equiv10^{-5}$ (upper panel) and data from the Heston model (cf. \eqref{eq.modelheston} and \eqref{eq.parametersheston}, lower panel). In each panel, the $x$-axis refers to different choices of the optimal constant $c^\star$ and the three curves represent different pre-average functions $\lambda_i$ ($\lambda_1$: solid line, $\lambda_3$: dotted line, $\lambda_4$: dashed line).}\label{fig:MSE}
\end{figure}

\begin{figure}[h!]
 \includegraphics[width=\textwidth]{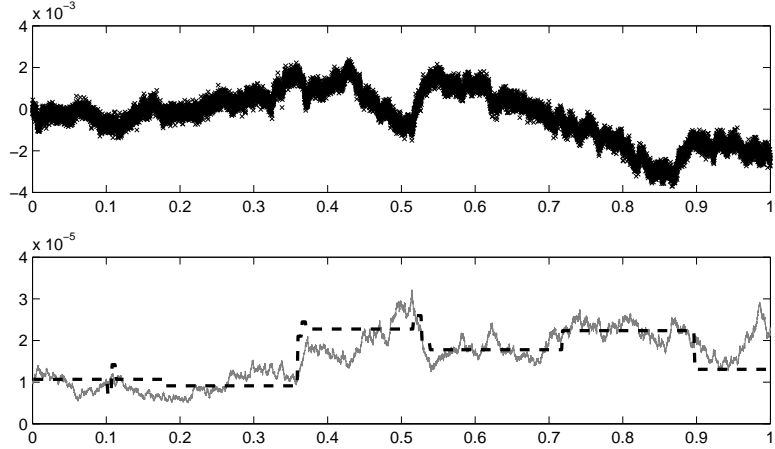}
\caption{Simulated data (Panel 1) coming from the Heston model with parameter as in \eqref{eq.parametersheston} for $n=15,000$ and true $\SNR\approx 15-20$, the true spot volatility function (solid line, Panel 2) and ASVE (dashed line, Panel 2).}\label{fig:reconstr}
\end{figure}

Figure \ref{fig:MSE} displays the results of the simulation study. In both panels, we choose $n=15,000$, $\SNR=20$ with constant $\tau$ and standard Gaussian white noise. Both display the empirical mean integrated squared error
\begin{equation}\MISE=\frac{1}{10,000}\sum_{i=1}^{10,000}\int^1_0(\hat\sigma^2_i(s)-\sigma^2_i(s))^2 ds\label{eq.MISE}\end{equation} based on 10,000 repetitions for $\lambda\in\{\lambda_1,\lambda_3,\lambda_4\}$ and different choices of the multiplicative constant $c$ (x-axis). In the upper panel, the data are generated with constant $\sigma$. In the lower panel, we simulate the latent log-price $X$ according to the Heston stochastic volatility model
\begin{align}
\begin{aligned}\label{eq.modelheston}
 dX_t &= -\frac 12 \sigma^2_t dt +\sigma_t dW_t, \\
 d\sigma^2_t&= \kappa \big(\theta-\sigma^2_t\big) dt+ \epsilon\sigma_t  d\widetilde W_t.
\end{aligned}
\end{align}
In this model, the Brownian motions $W$ and $\widetilde W$ are correlated, that is $dW_t d\widetilde W_t=\rho dt$ with $\rho \in [-1,1]$. It is not difficult to verify that $X$ is indeed a continuous semimartingale. The Heston model is commonly believed to describe stock market data quite well. It only depends on a few parameters which have a clear financial interpretation allowing in particular for leverage effects ($\rho<0$). For real data, estimates of the parameters in the Heston model have been carried out in different settings (see for instance Table 5.1 in \citet{Ploeg:2005}). 
For our simulations, we set 
\begin{align}\label{eq.parametersheston}
\rho=-2/3,\theta=10^{-5},\kappa=4,\epsilon=\sqrt{\kappa \theta}.                        
\end{align}
For these parameters, the volatility $\sigma^2$ typically takes values in $[2\cdot10^{-6},5\cdot10^{-5}]$, see also Figure \ref{fig:reconstr}.

From our simulation study, we find that in the Heston model, there is essentially no difference between the three candidate functions as long as $c^\star$ is chosen appropriately. However, $\lambda_4$ seems to produce the best estimators in terms of MISE, when the volatility function is constant. This is surprising, since from an asymptotic point of view, $\lambda_1$ is preferable. Our explanation is that non-asymptotically the boundary behavior of the pre-average function matters. Note that in contrast to $\lambda_i$, $i=1,3$, the function $\lambda_4$ vanishes at $0$ and $2$ and hence downweights observation at the end of the pre-average intervals $((i-2)/m,i/m]$.

Observe that the curves in the lower panel in Figure \ref{fig:MSE} are smoother than the ones in the upper panel. We explain this by the fact that the SNR is constant for deterministic $\sigma^2$ and varies in the Heston model. Thus, the randomness of the volatility has a smoothing effect and discretization effects become visible in the first case only.

In Figure \ref{fig:reconstr}, we illustrate the procedure for $\lambda=\lambda_4$ and $c=c^\star\cdot \widehat{\operatorname{SNR}}$. Here, $X$ follows again the Heston model with parameters given in \eqref{eq.parametersheston}. Observe that the stylized nature of the reconstruction only reflects the main features of $\sigma^2$.

\subsection{Jump detection}\label{sec:jumptest}

Note that out theoretical considerations are based on model \eqref{eq.mod}, that is assuming a continuous It\^o semimartingale as (log-) price process corrupted by additive noise. However, the continuity assumption in the model is often too strict in reality, since for example micro- or macroeconomic announcements may cause jumps in the price. The presence of such jumps is discussed in \citet{Ait-SahaliaJacod:2009}, \citet{BollerslevTodorov:2011}, and the references therein.

The most natural way to include a jump component into the model is to allow for non-continuous semimartingales. Estimation of the integrated volatility under microstructure noise and jumps has been considered for instance in \citet{PodolskijVetter:2009}. Eliminating jumps turns out to be much less difficult than taking microstructure noise into account. 

\begin{figure}[t!]
 \includegraphics[width=\textwidth]{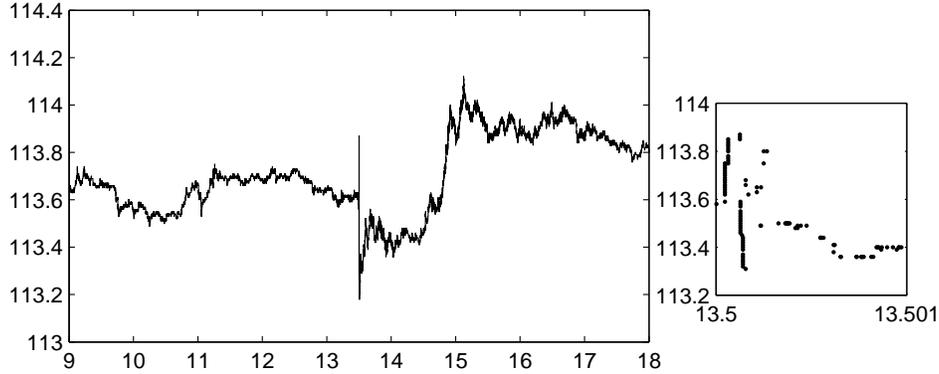}
\caption{FGBL data of November 2nd, 2007 and magnification of a small time interval around 1.30 p.m., where multiple consecutive jumps of the process occur.}\label{fig:multijumps}
\end{figure}

In order to correct for jumps, we adopt a rather practical point of view here. In fact, looking at financial data, relevant jumps seem to occur very irregularly. Occasionally, there are isolated jumps and quite rarely, jumps clustered over very short time intervals appear (cf. Figure \ref{fig:multijumps}). Therefore, our aim in this section is a hands-on approach to detect and to remove possible jumps as a pre-processing of the data. 

\begin{figure}[t!]
 \includegraphics[width=\textwidth]{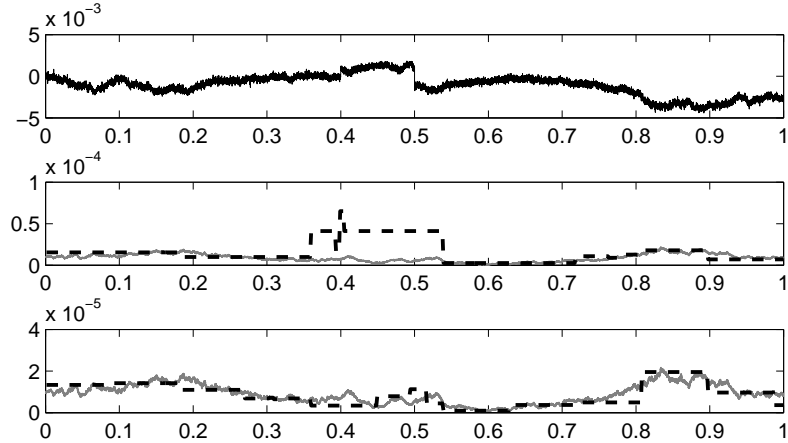}
\caption{Simulated data (Panel 1) coming from the Heston model with parameter choices given in \eqref{eq.parametersheston} for $n=15,000$ and true $\SNR\approx 15-20$ with two additional jumps at 0.4 and 0.5, the true spot volatility function (gray, solid line, Panel 2 and 3) and ASVE neglecting the presence of jumps (dashed line, Panel 2) and automatically finding and correcting the jumps (dashed line, Panel 3).}\label{fig:jumps}
\end{figure}

As usual, we model jumps as a c\`{a}dl\`{a}g jump process $(J_t)_t$. If jumps are present, ASVE will reconstruct the pointwise sum of the spot volatility plus the jump process $t\mapsto (J_t-J_{t-})^2$, where $J_{t-}$ denotes the left limit of $J$ at time point $t$. Note that $(J_t-J_{t-})^2$ is either zero or produces a spike depending on whether there is a jump at time point $t$ (cf. Figure \ref{fig:jumps}, Panel 1). In order to separate spot volatility and jump part, we apply the following method:

Let $m_1=\lfloor n^{3/4}\rfloor$ and $\lambda$ be a pre-average function. For $r=\frac n{m_1},\dots,n-\frac n{m_1}$, define
\begin{align}
Q_r:=\frac {m_1}n \sum_{j=r-\frac n{m_1}}^{r+\frac n{m_1}}\lambda\big(1+(j-r)\frac {m_1}n\big)Y_{j,n}.\label{eq.scan}
\end{align}
If there is no jump in $[r-\frac n{m_1},r+\frac n{m_1}]$, then $Q_r=O_P(n^{-1/8})$ (following the heuristic explanation in Subsection \ref{sec:heuristic}). Under the alternative, that is there is a jump with height $\Delta_r$ at $r/n$, we obtain $Q_r=O_P(\Delta_r)$. Note that by some CLT argument, $Q_r$ is approximately Gaussian distributed. Therefore, we may apply a procedure mimicking a local $t$-test:

\begin{enumerate}
 \item We partition the the set $\{Q_r:r=\tfrac{n}{m_1},\dots,n-\tfrac{n}{m_1}\}$ into blocks of length $n^{1/2}$.
 \item For each of these blocks, we compute the mean $\hat\mu$ and the standard deviation $\widehat{sd}$.
 \item For each $Q_r$ in a block, we compare $(Q_r-\hat\mu)/\widehat{sd}$ with a fixed threshold $t$. Here, simulations show that $t=2.81$ performs well.
\end{enumerate}

Afterwards, we reject those pre-averaged value $Z_{i,m}$, whose support intersects the support of a $Q_r$ rejected by the procedure. Those rejected values are replaced by the average of the nearest neighbors which are not rejected.

This procedure ensures that isolated jumps are detected. However, we often observe in real data that there are consecutive jumps within a short time period (cf. FGBL data of November 2nd, 2007 in Figure \ref{fig:multijumps} as an example). This may result in acceptance of the hypothesis that there is no jump, since a single jump might be not high enough in comparison to the estimated variance of $Q_r$. However, it is high enough to disrupt the performance of ASVE severely. To overcome this problem, we introduce a second test based on comparing increments of the observations directly which is more suitable to detect jump clusters. 

From our data sets, we find that the level of the microstructure noise, that is $\tau$, remains almost constant over a day. Thus, to explain the test, we might assume that $\tau$ is constant. Then,
\begin{align*}
 Y_{i,n}-Y_{i-1,n}=\tau(\eta_{i,n}-\eta_{i-1,n})+O_P(n^{-1/2})\approx \tau(\eta_{i,n}-\eta_{i-1,n}),
\end{align*}
if there is no jump. Secondly, we observe that the distribution of the noise is well-concentrated around zero. Thus, from a practical perspective, it is justified to assume that the tails of the microstructure noise are not heavier than that of a Gaussian random variable. If $(\eta_{i,n})$ would be i.i.d. standard normal, then using Corollary 2.1 in \citet{LiShao:2002}, we find  the following behavior regarding extreme values:
\begin{align*}\lim_{n\rightarrow \infty} \mathbb{P}(\max_{i=2,\ldots,n}(\eta_{i,n}-\eta_{i-1,n})^2 \leq 4\tau^2\log n)=1.\end{align*} 
Consequently, we identify the difference $Y_{i,n}-Y_{i-1,n}$ as due to a jump, if the squared increment exceeds $4\widehat \tau^2\log n$, where $\widehat \tau^2=(2n)^{-1}\sum_{i=2}^n (Y_{i,n}-Y_{i-1,n})^2$ is an estimator for $\tau^2$. Note that the latter procedure is much less powerful for isolated jumps than the first one, since it cannot detect jumps of size $o_P(\log n)$.

To illustrate the results, Figure \ref{fig:jumps} displays simulated data corrupted by two additional jumps at 0.4 and 0.5. ASVE without jump correction (Panel 2) incorporates a bump at the positions of the jumps. In contrast, pre-processing the data in a first step as outlined in this section yields a stable reconstruction (Panel 3).

A simulation study regarding the jump detection procedure is given in Section \ref{sec:jumpsim}.

\section{Simulations}\label{sec:sim}
\subsection{Stability}

\begin{table}[h]
\begin{center}
\renewcommand{\arraystretch}{1.5}
\begin{scriptsize}
\begin{tabular*}{\textwidth}{@{\extracolsep{\fill}}ccccc}
\hline
\multicolumn{2}{c}{std. of noise}	&$1/5,000$ 					&$3/5,000$					&$10/5,000$ \\
\hline
\multirow{2}{*}{MISE$\cdot10^{11}$}	&Gaussian	&$1.41$ ($3.28$)	&$2.39$ ($6.04$)	&$5.05$ ($14.34$)\\
			&uniform	&$1.40$ ($3.21$)	&$2.40$ ($6.10$)	&$5.08$ ($14.47$)\\
\hline
\multirow{2}{*}{rMISE}	&Gaussian	&$0.11$ ($0.20$)				&$0.19$ ($0.38$)				&$0.39$ ($0.94$)\\
			&uniform	&$0.12$ ($0.20$)				&$0.19$ ($0.38$)				&$0.40$ ($0.97$)\\
\hline
\end{tabular*} 
\end{scriptsize}
\caption{Stability under different distributions and levels of noise: MISE (upper row), rMISE (lower row), and respective 95\%-quantiles of the squared errors (in brackets) based on 10,000 repetitions.}
\label{tab:noise} 
\end{center}
\end{table}
\renewcommand{\arraystretch}{1}

To test the stability of ASVE, we simulate data for sample size $n=15,000$ and $X$ following the Heston SDE (cf. \eqref{eq.modelheston}) with parameters given in \eqref{eq.parametersheston}. To model the microstructure effects $(\epsilon_{i,n})_{i=1,\ldots,n}$, we consider Gaussian and uniform noise with standard deviations $x/5,000$ and $x\in \{1,3,10\}$. Here, a standard deviations of $1/5,000$ refers to a SNR of approximately $15$ and represents FGBL data best. We perform a simulation study with $10,000$ repetitions. Besides the mean integrated squared error (MISE, cf. \eqref{eq.MISE}), we investigated the behavior of the relative mean integrated squared error (rMISE), given by\[\rMISE=\frac{1}{10,000}\sum_{i=1}^{10,000}\frac{\int^1_0(\hat\sigma^2_i(s)-\sigma^2_i(s))^2 ds}{\int^1_0\sigma^4_i(s) ds},\] where $\hat\sigma^2_i$ and $\sigma_i^2$ refer to the estimated and the true volatility in run $i$. Throughout our simulations, we use Haar wavelets and $\lambda_4$ as a pre-average function. Following 
Section 
\ref{sec:opttuning}, we set $c=0.3\cdot \widehat{\operatorname{SNR}}$. The results and empirical 95\%-quantiles are displayed in Table \ref{tab:noise}. We observe that the outcome is essentially not affected by the distribution. In contrast, the SNR has a large impact on the performance (recall that $\sigma^2\approx 10^{-5}$). The bad performance of 
the estimator for the largest standard deviation can be explained by the choice of $m$, which is inversely proportional to the noise level. In fact the optimal oracle would be $m_{\text{oracle}}=0.3\cdot SNR \sqrt{n}\approx 55$. Thus, regarding the problem as a $\chi_1^2$-regression problem (cf. Section \ref{sec:heuristic}), we have to estimate $\sigma^2$ based on 55 observations, which is a quite difficult task.

\subsection{Robustness}\label{sec:jumpsim}
As discussed in Section \ref{sec:jumptest}, there are two major model violations one has to take into account for real data, namely rounding effects and jumps. In a simulation study, we investigate the robustness of ASVE with and without jump detection given data with rounding errors and jumps. The process $X$ is generated from the Heston model \eqref{eq.modelheston} with parameters as in \eqref{eq.parametersheston}. This ensures that the $\SNR$ lays most of the time between $15$ and $20$. Mimicking real FGBL prices, the sample size or the number of trades per day is $n=15,000$. Here, rounding means rounding the corresponding price ($110\exp(Y_{i/n})$) up to the two decimal places, and afterwards transforming back via $\log(\tfrac{\cdot}{110})$, that is rounding to full basis points of the price and is not to be confused with rounding of the log-price. Notice that FGBL prices are most of the time in the range between $100$ and $120$. Therefore, $110$ is a reasonable starting value (cf. also the upper panel 
in Figure \ref{fig:intro}). The jump process is simulated as a compound 
Poisson process with constant intensity $1/3$ and jump size distribution $\mathcal{N}(0,10^{-6})$.

\begin{table}[h]
\centering
\begin{tabular*}{\textwidth}{@{\extracolsep{\fill}}ccccc}
	\hline
	& pure & rounded & with jumps & with jumps, rounded \\
	\hline
	without & \multirow{2}{*}{$1.41\cdot10^{-11}$} & \multirow{2}{*}{$1.41\cdot10^{-11}$} & \multirow{2}{*}{$12.64\cdot10^{-11}$} & \multirow{2}{*}{$ 12.86\cdot10^{-11}$} \\
jump detection\\
\hline
	with  & \multirow{2}{*}{$1.68\cdot10^{-11}$} &\multirow{2}{*}{$1.69\cdot10^{-11}$} & \multirow{2}{*}{$1.69\cdot10^{-11}$} & \multirow{2}{*}{$1.70\cdot10^{-11}$} \\
jump detection\\
	\hline
\end{tabular*} 
\caption{Robustness. Simulation results for the MISE for data generated from the Heston model with additional rounding and jumps for ASVE with and without jump detection.}
\label{tab:robust} 
\end{table}

The resulting empirical mean integrated squared errors (MISE) computed on the basis of 10,000 repetitions are displayed in Table \ref{tab:robust}. Obviously, jumps have a huge influence on ASVE, while rounding effects are negligible (at least regarding the FGBL data sets in Section \ref{sec:real}). We observe that the bad impact of the jumps is reduced almost completely by the pre-processing of the data.

\section{Time schemes}\label{sec:ticktime}

It has been noticed in the econometrics literature that an increase in volatility might be due to different reasons. One explanation would be that there are larger price changes. Alternatively, the volatility will of course also increase if price changes are of the same size and only the number of trades per time interval goes up (cf. for example \citet{EderingtonLee:1995}, Section IV.B). Disentangling the different explanations is quite difficult without an underlying mathematical concept. Nevertheless, determining the source of an increase in volatility is clearly of importance.

A more rigorous treatment of this problem leads to the definition of different notions of time (for instance in \citet{DahlhausNeddermeyer:2013}). Here, we investigate the most prominent examples: real time and tick time (sometimes also referred to as clock time and transaction time). 

Volatility in real time is appealing as it seems very intuitive. In tick time successive ticks are treated as one time unit. By definition, this time scheme does not depend on the speed at which successive trades occur. Consequently, volatility in tick time is independent of the trading intensity and hence measures the volatility of the price changes only. As the trading speed can be estimated directly from the ticks, we argue in this section that tick time volatility is the more natural object. A drawback of tick times is that there is no straightforward extension of the concept to multivariate processes.

Let us clarify the connection between both time schemes in more detail. 
Denote by $t_i, \ i=1,\ldots,n$ the ordered ($t_0<t_1<t_2<\ldots<t_n$) sample of trading times. Then, for $i<j$ the time between $t_i$ and $t_j$ equals $\tfrac{j-i}n$ time units in tick time and $t_j-t_i$ time units in real time. With this notation, the tick time model is given by
\begin{align}
  Y^T_{i,n}=X_{t_i}+\epsilon_{i,n}, \quad i=1,\ldots,n.
  \label{eq.mod_T}
\end{align}
Inspired by the classical high-frequency framework, we think about the trading times as an array, that is $t_i=t_{i,n}$, where the sampling rate gets finer for increasing $n$. Define the trading intensity $\nu$ at time $t$ as
\begin{align}
  \nu(t)=\lim_{n\rightarrow\infty}\frac{\tfrac 1n \sum_{i=1}^n \mathbb{I}_{[t-\delta_n, t+\delta_n]}(t_i)}{2\delta_n}(t_n-t_0),
  \label{eq.nu_t_def}
\end{align}
provided this limit exists and is unique for any sequence $\delta_n\rightarrow 0$ and $\delta_n n \rightarrow \infty$. 

As an example consider the following toy model: Assume that $\sigma$ is deterministic and there exists a deterministic, differentiable function $h:[0,1]\rightarrow[0,1]$ with  $h(i/n)=t_{i,n}$ (in particular this implies that $h$ is strictly monotone). Note that in this setting, $\nu$ is deterministic as well and given by the derivative of $h^{-1}$. 

Let $\sigma_{RT}$ denote the original (real time) volatility. Recall that under tick time, we consider successive trading times as equidistant. Therefore, the tick time volatility $\sigma_{TT}$ satisfies for all $i=1,\ldots,n$
\begin{align*}
\int^{i/n}_0\sigma_{TT}(h(s))dW_s= \int^{h(i/n)}_0\sigma_{RT}(s)dW_s=_\mathcal{L}\int^{i/n}_0 \sqrt{h'(s)}\sigma_{RT}(h(s))dW_s
\end{align*}
in law. Thus, the first and the latter integrand are (roughly) equal, that is $\sigma_{TT}^2(h(s))=h'(s)\sigma^2_{RT}(h(s))$. Rewriting this, we obtain \begin{align}\label{eq.ticktimerealtime}\nu \sigma_{TT}^2=\sigma_{RT}^2,\end{align} cf. also \citet{DahlhausNeddermeyer:2013}, Section 4. This formula clarifies the connection between tick time and real time volatility. 

Estimating the real time volatility directly from tick data, we have to construct artificial observations by recording the price each 10th second, for example. This method leads to a loss of information if there are many ticks in one time interval.

Notice that nonparametric estimation of the trading intensity $\nu$ is standard using for example \eqref{eq.nu_t_def} together with a proper choice of the bandwidth $\delta_n$. In view of formula \eqref{eq.ticktimerealtime}, it seems therefore more natural to estimate the real time spot volatility as product of $\widehat \sigma_{TT}^2$ and an estimator of $\nu$. In a simulation study, we estimated the real time volatility via its product representation for Euro-BUND Futures on all days in 2007 (for a description of the data, cf. also Section \ref{sec:real}). We use Haar wavelets and hence obtain piecewise constant reconstructions. As a measure for the oscillation behavior of the volatility, we take the sum of squared jump sizes of the reconstructions for every of these days. In average, for tick time spot volatility this gives $9.68\cdot10^{-11}$ per day, while for real time volatility the corresponding value is $1.98\cdot10^{-10}$. This gives some evidence that the tick time volatility is much smoother than 
its real 
time counterpart. 

As a surprising fact, formula \eqref{eq.ticktimerealtime} shows that even rates of convergence for estimation of $\sigma_{RT}^2$ can be much faster than the minimax rates provided $\sigma_{TT}^2$ is sufficiently smooth. To give an example, assume that $\sigma_{TT}$ is constant and $\nu$ has H\"older continuity $\beta>1/2$. In this case $\nu$ can be estimated with the classical nonparametric rate $n^{-\beta/(2\beta+1)}\ll n^{-1/4}$. Consequently, $\sigma_{RT}^2$ has also H\"older index $\beta$. The rate for estimation of $\sigma_{RT}^2$ is $n^{-1/4}$ which converges faster to zero than the minimax rate $n^{-\beta/(4\beta+2)}$ (for a derivation of minimax rates see \citet{MunkSchmidt-Hieber:2010a} and Hoffmann et al.).

To summarize, the tick time volatility is the quantity of interest measuring the volatility of the price changes. Furthermore, the real time volatility can easily be estimated via \eqref{eq.ticktimerealtime}. For these reasons, we restrict ourselves throughout the following to estimation of spot volatility in tick time.

\section{Spot volatility of Euro-BUND Futures}\label{sec:real}

We analyze the spot volatility of Euro-BUND Futures (FGBL) using tick data from Eurex data\-base. The underlying is a 100,000 Euro debt security of the German Federal Government with coupon rate 6\% and maturity $8.5-10.5$ years. The price is given in percentage of the par value. The tick times are recorded with precision of 10 milliseconds. The minimum price change is $0.01 \%$ (one basis point), corresponding to 10 Euro, which is comparably large. The number of trades per day varies among 10,000 and 30,000. Observations which are not due to trading are removed from the sample. If there are different FGBL contracts at a time referring to different expiration days, we only consider these belonging to the next possible date. Trading takes places from 8:00 a.m. until 7:00 p.m. Central European Time (CET). For the reconstructions, we restrict ourselves to observations within the time span 9 a.m. to 6 p.m. CET. Outside this period, trading is normally too slow to make use of a high-frequency setting.

\begin{figure}
 \includegraphics[width=\textwidth]{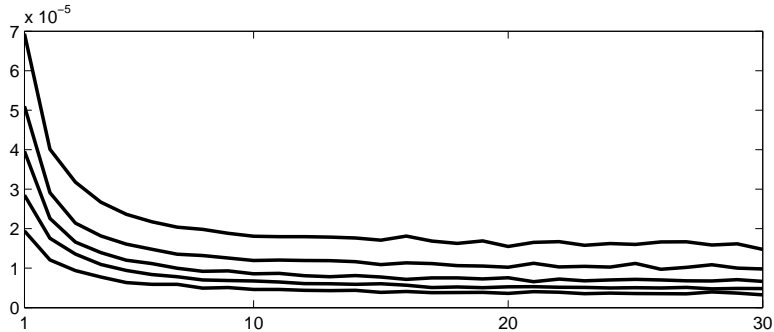}
 \caption{Realized volatilities of FGBL data from June 4th to June 8th, 2007 for different subsampling frequencies.}\label{fig:RV_sub}
\end{figure}

During business hours, FGBL prices fit well as an example for high-frequency data. On the one hand, trading is very liquid due to low transaction costs and high trading volume. In average, the holding period is less than two days (cf. \citet{Dorfleitner:2004}, Figure 4). On the other hand, microstructure effects are present and simple quadratic variation techniques fail as indicated in Figure \ref{fig:RV_sub}. In this plot (often referred to as signature plot), we investigate how the (integrated) realized volatilities behaves if we consider subsamples of the data with different subsampling frequencies. We observe a rapid increase on small frequencies, that is if more and more data are included. This indicates that microstructure effects have to be taken into account.

In the following, we illustrate the effect of macroeconomic events with unanticipated outcome on spot volatility. As they cause uncertainty, one expects an increase in volatility once they appear. There has been a large body of literature in economics devoted to this subject. Nevertheless, up to now, there seems to be no general consensus quantifying how much the volatility is affected by public announcements. \citet{EderingtonLee:1993} and \citet{EderingtonLee:1995} claim that volatility is substantially higher for a few minutes after the announcement and is still visible in the data for several hours. They also find evidence that volatility is slightly elevated for some minutes before an announcement. They conclude that macroeconomic announcements are the driving force for volatility. In contrast, in the seminal paper \citet{AndersenBollerslev:1998} daily volatility patterns are found to explain most of the spot volatility behavior, while public announcements have only a secondary effect on overall 
volatility. In a recent study, \citet{LundeZebedee:2009} focus on the effects of US monetary policy events on volatility of US equity prices. In accordance with previous work, they conclude that there are spikes in the volatility around macroeconomic announcements, lasting for approximately 15 minutes. In \citet{JansendeHaan:2006} effects of certain European Central Bank (ECB) announcements on price changes and volatility are studied. Although these papers deal with volatility on relatively short time intervals, none of them accounts for microstructure effects.

\begin{figure}[h!]
 \includegraphics[width=\textwidth]{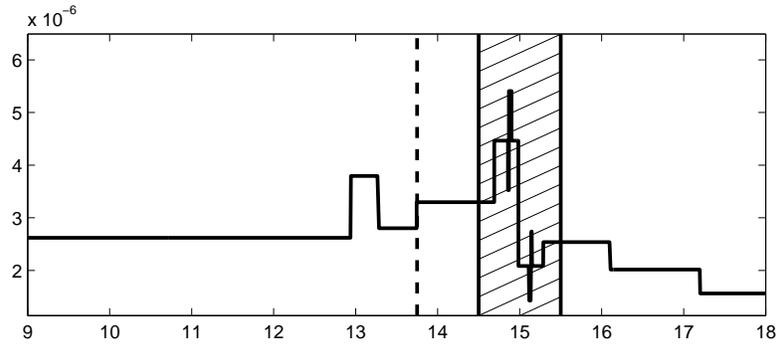}
\caption{ASVE for May 10th, 2007. Period of the ECB press conferences is hatched and announcement of not changing the key interest rate is represented by the dashed line).}\label{fig:keyinterest}
\end{figure}

To illustrate our method, the twelve days in 2007 (one per month) with an official ECB press conference related to possible changes in key interest rates are studied. During these meetings hold jointly by the president and the vice-president of the European Central Bank, announcements about ECB-policy are made. In \citet{JansendeHaan:2006}, press conferences are excluded from the study, but they are very appealing because on the one hand, key interest rates are of major economic importance especially for government bonds like Euro-BUND futures, and on the other hand, the announcement procedure is highly standardized. In fact, on every of the studied dates the decision of the ECB Governing Council on the key interest rates was released on 1.45 p.m. followed by the official press conference starting at 2.30 p.m. and lasting for exactly an hour. The press conference consists of two parts starting with an introductory statement by the ECB president. In a second part, the president and vice-president answer 
questions of journalists. On every of these events, between 20 
and 62 financial analysts are asked in advance to predict possible changes in the key interest rate. Based on these estimates a sample standard deviation is computed which is available at Bloomberg. In the following, we refer to this quantity as {\it market uncertainty}. 

In Figure \ref{fig:keyinterest}, ASVE for May 10th, 2007 is displayed. The dashed line represents the time of the announcement, the hatched region refers to the time period of the press conference. On this day, the reconstruction displays an increase in volatility around the time of the announcement. Furthermore, we observe a higher fluctuation during the press conference. A more thorough analysis is done in Table \ref{tab:ecb}: We observe a slight increase of the spot volatility on most of the considered days in view of average, maximum and total
variation (which reflects the volatility of the volatility). On days, where the market uncertainty was nonzero, this effect is even enhanced. Notice that the integral and TV figures are normalized by the length of the time interval to make them comparable. The results confirm the influence of macroeconomic events on volatility.

\begin{table}[h]
\centering
\begin{tabular*}{\textwidth}{@{\extracolsep{\fill}}l|c|ccc|ccc}
\hline
\multirow{2}{*}{Day}&    Market&    \multicolumn{3}{c|}{13.40 pm - 13.50 pm} &            \multicolumn{3}{c}{13.45 pm - 15.30 pm}\\
&uncertainty&                            $\int\hat\sigma^2$&$\max\hat\sigma^2$&$\TV\hat\sigma^2$&    $\int\hat\sigma^2$&$\max \hat\sigma^2$&$\TV \hat\sigma^2$\\
\hline
Jan-11&            0&                $0.459$&$0.459$&$0$&                        $0.435$&$0.518$&$0.168$\\
Feb-08&            0&                $0.541$&$0.541$&$0$&                        $0.509$&$0.979$&$1.485$\\
Mar-08&            0&                $0.490$&$0.490$&$0$&                        $0.497$&$0.643$&$0.685$\\
Apr-12&            0&                $0.274$&$0.331$&$1.222$&                    $0.374$&$0.698$&$0.472$\\
May-10&            0&                $0.318$&$0.330$&$0.298$&                    $0.323$&$0.541$&$0.594$\\
Jun-06&            0&                $0.191$&$0.191$&$0$&                        $0.495$&$0.677$&$0.455$\\
Jul-05&            0&                $0.490$&$0.587$&$0.772$&                    $0.683$&$1.315$&$1.045$\\
Aug-02&            0.05&                $0.745$&$1.286$&$8.673$&                    $1.176$&$5.749$&$7.075$\\
Sep-06&            0.1&                $0.906$&$0.906$&$0$&                        $0.969$&$2.862$&$5.626$\\
Oct-04&            0.03&                $0.621$&$0.621$&$0$&                        $0.701$&$1.181$&$0.936$\\
Nov-08&            0&                $0.869$&$0.869$&$0$&                        $1.020$&$1.337$&$0.480$\\
Dec-06&            0&                $1.119$&$1.119$&$0$&                        $0.958$&$2.545$&$3.150$\\
\hline
\multicolumn{2}{c|}{average of days above}&        $0.585$&$0.644$&$0.914$&                    $0.678$&$1.587$&$1.848$\\
\hline
\multicolumn{2}{c|}{average of all days}&        $0.515$&$0.551$&$0.621$&                    $0.552$&$1.225$&$1.328$\\
\multicolumn{2}{c|}{$90\%$-quantile all days}&        $0.906$&$0.960$&$0.661$&                    $0.984$&$2.051$&$2.609$\\
\hline
\end{tabular*}
\caption{Features (average, maximum, and total variation) of ASVE for days with ECB press conferences on key interest rates. The second column is an estimate of market uncertainty. Integrated volatility and total variation are normalized by the length of the time interval. All entries related to volatility are multiplied by $10^5$.}
\label{tab:ecb} 
\end{table}

\section{Generalization to Spot Covolatility Estimation}\label{sec:multi}

So far, we considered one-dimensional processes only. As for example in portfolio management, one might more generally be interested in the spot covariance matrix of multi-dimensional (and even very high-dimensional) price processes. There has been a lot of recent interest in this direction. The main additional difficulty is to deal with non-synchronous observations. Synchronization schemes in the context of estimation of the integrated covolatility (the multi-dimensional extension of the integrated volatility) were proposed in \citet{HayashiYoshida:2005}, \citet{Ait-SahaliaFanXiu:2010}, \citet{ChristensenKinnebrockPodolskij:2010}, \citet{Barndorff-NielsenHansenLundeShephard:2011}, \citet{Zhang:2011}, and \citet{Bibinger:2011}, among others.

As an outlook, we shortly point out how to construct an estimator of the spot covolatility function $\cov$ given synchronous data, that is the covariance function of two price processes observed at the same time points. To the best of our knowledge, nonparametric estimation of the spot covolatility under microstructure noise has not been treated so far. For simplicity, we restrict ourselves to the bivariate case. In principle, this estimator can be combined in a second step with any of the synchronization schemes mentioned above. 

Assume that we observe two processes
\begin{align}
  Y^{(1)}_{i,n}=X_{i/n}^{(1)} +\epsilon_{i,n}^{(1)}, \quad
  Y^{(2)}_{i,n}=X_{i/n}^{(2)} +\epsilon_{i,n}^{(2)}, \quad i=1,\ldots,n,\label{eq.model.cov}
\end{align}
where $dX^{(1)}_t=\sigma^{(1)}_tdW^{(1)}_t$ and $dX^{(2)}_t=\sigma^{(2)}_tdW^{(2)}_t$ are two It\^o martingales with driving Brownian motions $W^{(1)},W^{(2)}$, and $\epsilon^{(1)},\epsilon^{(2)}$ are two independent noise processes each defined analogously to \eqref{eq.noise}. We assume that the spot covolatility function of $X^{(1)}$ and $X^{(2)}$ is given by $\cov_t\, dt=\Cov(dX^{(1)}_t,dX^{(2)}_t)$.

\begin{figure}[h!]
 \includegraphics[width=\textwidth]{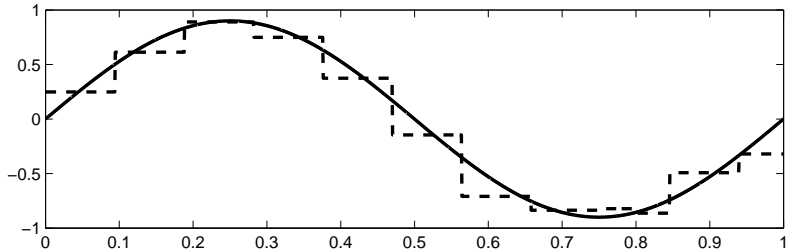}
\caption{Reconstruction (dashed) and true covolatility function (solid) for data following model \eqref{eq.model.cov} ($n=15,000$), constant volatilities $\sigma_1,\sigma_2$, and i.i.d. centered Gaussian noise with $\SNR=\tfrac{\sqrt{\langle|\cov|,1\rangle}}{\tau}=20$}\label{fig:cov}
\end{figure}

For $i=2,\dots,m$ and $q=1,2$, let $\overline{Y}_{i,m}^{(q)}$ be as defined in \eqref{eq.ovYdef}. Then, the wavelet coefficients of the spot covolatility are estimated via
\[\widehat{\langle g, \cov\rangle} :=\sum_{i=2}^m g\big(\tfrac{i-1}{m}\big)\overline{Y}^{(1)}_{i,m}\overline{Y}^{(2)}_{i,m}.\]
where again $g\in \{ \phi_{j_0,k}, \psi_{j,k} \}$. Since the noise processes $\epsilon^{(1)},\epsilon^{(2)}$ are independent, no bias correction is necessary.

For illustration, Figure \ref{fig:cov} shows the reconstruction of the covolatility function of a realization in model \eqref{eq.model.cov} using the same thresholding procedure and parameter choices as for ASVE.

\section*{Acknowledgment}
We thank the CRC 649 "Economic Risk" for providing us with access to Eurex database. Parts of this work are taken from the PhD thesis \citet{Schmidt-Hieber:2010}. We are grateful to Marc Hoffmann, Markus Rei\ss, and Markus Bibinger for many interesting discussions.

\begin{supplement}[id=suppA]
  \sname{Supplement}
  \stitle{Supplement to ''Spot volatility estimation for high-frequency data: adaptive estimation in practice``}
  \slink[doi]{ }
  \sdatatype{}
  \sdescription{In this supplement, we provide the proof of Lemma \ref{lem:mse}.}
\end{supplement}

\bibliographystyle{imsart-nameyear.bst}
\bibliography{References}
\newpage

\begin{frontmatter}
\title{Supplement to ``Spot volatility estimation for high-frequency data: adaptive estimation in practice''}
\runtitle{Spot volatility estimation in practice}

\begin{aug}
\author{\fnms{Till} \snm{Sabel}\corref{}\thanksref{m1,m3}\ead[label=e1]{tsabel@uni-goettingen.de}},
\author{\fnms{Johannes} \snm{Schmidt-Hieber}\thanksref{m2,m3}\ead[label=e2]{Johannes.Schmidt.Hieber@ensae.fr}}
\and
\author{\fnms{Axel} \snm{Munk}\thanksref{m1,m3}\ead[label=e3]{amunk1@gwdg.de}}

\thankstext{m3}{The research of the authors was supported by DFG/SNF-Grant FOR 916.}
\thankstext{m2}{The research of J. Schmidt-Hieber was funded by DFG postdoctoral fellowship SCHM 2807/1-1.}

\runauthor{T. Sabel et al.}

\affiliation{Georg-August-Universit\"at G\"ottingen\thanksmark{m1} and CREST-ENSAE\thanksmark{m2}}

\address{Institut f\"ur mathematische Stochastik\\Georg-August-Universit\"at G\"ottingen\\Goldschmidtstr. 7\\ 37077 G\"ottingen\\ Germany\\
\printead{e1}\\
\phantom{E-mail:\ }\printead*{e3}}

\address{\'Ecole Nationale de la Statistique\\ et de l'Administration \'Economique\\ Centre de Recherche\\ en \'Economie et Statistique\\ 3 Avenue Pierre Larousse\\ 92245 Malakoff\\ France\\
\printead{e2}}
\end{aug}

\begin{abstract}
In this supplement, we recall the proof of Lemma 3.1 in ``Spot volatility estimation for high-frequency data: adaptive estimation in practice'' as it is given in \citet{Schmidt-Hieber:2010}, Lemma 6, p. 65.
\end{abstract}

\end{frontmatter}
\begin{appendix}
\section{Proof of Lemma 3.1}
To keep notation simple, we use the following quantities in the spirit of the definitions of Section 2.3: For any process $(A_{i,n})\in\{(Y_{i,n}),(\epsilon_{i,n}),(X)_{i,n}\}$, define
\begin{align*}
 \overline A_{i,m}=\overline A_{i,m}(\lambda):=\frac mn \sum_{\frac jn\in[\frac{i-2}m,\frac{i}m]}\lambda\big(m\tfrac
jn-(i-2)\big)A_{j,n}.
\end{align*}
\begin{align*}
\mathfrak{b}(A)_{i,m}=\mathfrak{b}(\lambda, A_\cdot)_{i,m} 
:=  &\frac{m^2}{2n^2} \sum_{\frac jn\in[\frac{i-2}m,\frac im]}\lambda^2\big(m\tfrac jn-(i-2)\big)\big(A_{j,n}-A_{j-1,n}\big)^2.
\end{align*}

Further, recall that our estimator for the integrated volatility is given by $\widehat{\langle 1,\sigma^2\rangle}=\sum_{i=2}^m\overline Y_{i,m}^2-\mathfrak{b}(Y)_{i,m}$.

To prove the lemma, let us first show that the bias is of smaller order than $n^{-1/4}$. In fact, note that $\E\big[ \ \overline{Y}_{i,m}^2\big]= \E\big[ \ \overline{X}_{i,m}^2\big]+\E\big[ \overline{\epsilon}_{i,m}^2\big].$
Clearly, one can bound 
\begin{align*}
    \Big|\E\big[\overline{\epsilon}_{i,m}^2\big]-\E\big[\mathfrak{b}(\lambda,Y)_{i,m}\big]\Big|=O(\tfrac{1}{n}).
\end{align*}
Further, Lipschitz continuity of $\lambda$ together with a Riemann approximation argument gives us
\begin{align*}
    \big|\E\big[ \ \overline{X}_{i,m}^2\big]-\tfrac{\sigma^2}m\big|=& \Big|\tfrac{\sigma^2}m\int^2_0\int^2_0\lambda(s)\lambda(t)(s\wedge t)dtds-\tfrac{\sigma^2}m\Big|+O(\tfrac{1}{n})=O(\tfrac{1}{n}).
\end{align*}
Here, the last equation is due to partial integration and the definition of a pre-average function (cf. Definition 2.1). Since both approximations are uniformly in $i$, this shows that the bias is of order $O(n^{-1/2}).$

For the asymptotic variance, first observe that $\Var(\sum_{i=2}^m\mathfrak{b}(\lambda,Y)_{i,m})=o(n^{-1/2}).$ Hence, \[\Var(\widehat{\langle 1,\sigma^2\rangle})= \Var(\sum_{i=2}^m \overline{Y}_{i,m}^2)+o\Big(n^{-1/4}\big(\Var(\sum_{i=2}^m \overline{Y}_{i,m}^2)\big)^{1/2}+n^{-1/2}\Big),\] by Cauchy-Schwarz inequality. Recall that for centered Gaussian random variables $U,V$, $\Cov(U^2,V^2)=2(\Cov(U,V))^2$. Therefore, it suffices to compute $\Cov(\overline{Y}_{i,m},\overline{Y}_{k,m})=\E[\overline{Y}_{i,m} \overline{Y}_{k,m}]$. 

By the same arguments as above, that is Riemann summation and partial integration, we find
\begin{align*}
  \E\Big[\Big|\overline{X}_{i,m} \overline{X}_{k,m}-\int_0^1\Lambda(ms-(i-2))dX_s\int_0^1\Lambda(ms-(k-2))dX_s\Big|\Big]\lesssim n^{-1}.
\end{align*}
Therefore,
\begin{align*}
  \E\big[\overline{X}_{i,m} \overline{X}_{k,m}\big]=\sigma^2 \int_0^1 \Lambda(ms-(i-2))\Lambda(ms-(k-2))ds+O(n^{-1}),
\end{align*}
where the last two arguments hold uniformly in $i, k.$

In order to calculate $\E[\overline{Y}_{i,m}\overline{Y}_{k,m}],$ we must treat three different cases, $|i-k|\geq 2, \ |i-k|=1$ and $i=k,$ denoted by $I, \ II$ and $III.$

\subsection*{I:} In this case $(\tfrac{i-2}m,\tfrac im]$ and $(\tfrac{k-2}m,\tfrac km]$ do not overlap. By the equalities above, it follows $\Cov(\overline{Y}_{i,m},\overline{Y}_{k,m})=O(n^{-1}).$

\subsection*{II:} Without loss of generality, we set $k=i+1.$ Then, we obtain
\begin{align*}
  &\Cov(\overline{Y}_{i,m},\overline{Y}_{i+1,m})
  =
  \E\big[\overline{X}_{i,m}\overline{X}_{i+1,m}\big]
  +\E\big[\overline{\epsilon}_{i,m}\overline{\epsilon}_{i+1,m}\big] \\
  =&
  \sigma^2\int_0^1 \Lambda(ms-(i-2))\Lambda(ms-(i-1)) ds+O(n^{-1}) \\
  &+\tau^2 \frac{m^2}{n^2}\sum_{\tfrac jn \in \big(\tfrac{i-2}m,\tfrac im\big]}
   \lambda(m\tfrac jn-(i-2)) \lambda(m\tfrac jn-(i-1)) \\
  =& \frac{\sigma^2}m \int_0^1\Lambda(u)\Lambda(1+u) du
  +\tau^2 \frac mn \int_0^1  \lambda(u) \lambda(1+u) du
  +O(n^{-1}),
\end{align*}
where the last inequality can be verified by Riemann summation. Noting that $ \lambda$ is a pre-average function, we obtain $\lambda(1+u)=-\lambda(1-u)$ and 
\begin{align*}
&\Cov(\overline{Y}_{i,m},\overline{Y}_{i+1,m})\\= &\frac{\sigma^2}m \int_0^1\Lambda(u)\Lambda(1-u) du
  - \frac{\tau^2m}n \int_0^1  \lambda(u) \lambda(1-u) du
  +O(n^{-1}).
\end{align*}

\subsection*{III:} It can be shown by redoing the arguments in $II$ that
\begin{align*}
  &\Var(\overline{Y}_{i,m})=\Var(\overline{X}_{i,m})+\Var(\overline{\epsilon}_{i,m})\\
  = &\frac{\sigma^2}m \int_0^2\Lambda^2(u) du
  +\tau^2 \frac mn \int_0^2  \lambda^2(u) du
  +O(n^{-1}).
\end{align*}

Note that $\|\Lambda\|_{L^2[0,2]}=1.$ Since the above results hold uniformly in $i,k,$ it follows directly that
\begin{align*}
  &\Var(\sum_{i=2}^m \overline{Y}_{i,m}^2)\\
= &\sum_{i,k=2, \ |i-k|\geq 2}^m
   2\big(\Cov(\overline{Y}_{i,m}, \overline{Y}_{k,m})\big)^2\\
&+2\sum_{i=2}^{m-1} 2\big(\Cov(\overline{Y}_{i,m}, \overline{Y}_{i+1,m})\big)^2
  +\sum_{i=2}^m 2\big(\Var(\overline{Y}_{i,m})\big)^2 \\
  =& O(n^{-1})
  +4\Big(\frac{\sigma^2}{\sqrt{c}}\int_0^1\Lambda(u)\Lambda(1-u)du
  -\tau^2c^{3/2}\int_0^1 \lambda(u) \lambda(1-u) du\Big)^2n^{-1/2}
  \\
  &+2\Big(\frac{\sigma^2}{\sqrt{c}}
  +2\tau^2c^{3/2}\| \lambda\|_{L^2[0,1]}^2\Big)^2n^{-1/2}.\qquad\square
\end{align*}
\end{appendix}

\end{document}